# Zero-Dimensional Superconducting Fluctuations and Fluctuating Diamagnetism in Lead Nanoparticles

**E.Bernardi, A. Lascialfari* and A. Rigamonti**

*Department of Physics "A. Volta" and Unità CNISM-CNR, University of Pavia, Via Bassi 6, I-27100, Pavia (Italy)*
** currently at Institute of General Physiology and Biological Chemistry, University of Milano, Via Trentacoste 2, I-20134 Milano (italy)*

**L.Romanò**

*Department of Physics and Unità CNISM-CNR, University of Parma, Parco Area delle Scienze 7A, I-43100, Parma (Italy)*

**V. Iannotti, G. Ausanio**

*Coherentia CNR-INFM, Physical Science Department, "Federico II" University, Piazzale V. Tecchio 80 , I-80125, Napoli (Italy)*

**C. Luponio**

*Department of Materials Engineering and Production,"Federico II" University, Piazzale V. Tecchio 80, I-80125, Napoli (Italy)*

## Abstract

High resolution SQUID magnetization measurements in lead nanoparticles are used to study the fluctuating diamagnetism in zero-dimensional condition, namely for particle size d lesser than the coherence length $\xi$. The diamagnetic magnetization $M_{dia}$ (H, T= const) as a function of the field H at constant temperature is reported in the critical region and compared with the behaviour in the temperature range where the first-order fluctuation correction is expected to hold.

The magnetization curves are analysed in the framework of exact fluctuation theories based on the Ginzburg-Landau functional for $\xi >> d$. The role of the upturn field $H_{up}$ where $M_{dia}$ reverses the field dependence is discussed and its relevance for the study of the fluctuating diamagnetism, particularly in the critical region where the first-order fluctuation correction breaks down, is pointed out. The size and temperature dependence of $H_{up}$ is theoretically derived and compared to the experimental data.

The relevance of the magnetization curves for non-evanescent field and of the upturn field for the study of the fluctuating diamagnetism above the superconducting transition temperature is emphasized.

## Introduction

Superconducting fluctuations (SF) and precursor effects occurring above the superconducting transition temperature $T_c$ have attracted a great deal of interest along the last decades. The effects of SF on transport properties have been studied and analysed in the light of advanced theoretical descriptions[1]. Particular interest involves the effect of an external magnetic field. While the field is a necessary tool to carry out significant types of measurements, one has to expect that it tends to suppress the fluctuating Cooper pairs. A variety of interplay phenomena in different temperature ranges and involving the critical superconducting fields, dimensionality or impurities are actually reflected in the dependence of the SF and of the fluctuating diamagnetism (FD) from the external field H. On the other hand the effect of H on SF is an intricate issue and in some cases its real detectability appears questionable. This, for instance, happens for the field dependence of the spin susceptibility controlling the NMR spin-lattice relaxation[2] or for the diamagnetic susceptibility $\chi_{dia}$ related to FD[3].

Relevant advantages in the study of the field-dependence of SF occur for grains of size d smaller than the coherence length $\xi$. In fact in these samples, besides the enhancement of the fluctuations, one has an enlargement of the critical region, that can conveniently be investigated experimentally. Meantime only spatially uniform fluctuations of the order parameter $\Psi(\mathbf{r})$ can be considered and thus the Ginzburg-Landau (GL) free energy functional does no longer require the expansion of $\Psi(\mathbf{r})$ in Fourier series and exact solutions for any field $H \ll H_c$ become possible. Understanding the field dependence of SF for grain size $d \ll \xi$ is useful for the FD in bulk superconductors also. In fact, the dominant contribution to the diamagnetic susceptibility $\chi_{dia}$ for $T \rightarrow T_c^+$ originates from metastable superconducting "droplets" of diameter of the order of $\xi(T)$. Then the isothermal magnetization curves $M_{dia}$ ( T= const, H), at least qualitatively can be discussed by assuming for the fluctuating droplets the zero-dimensional (0D) condition and valuable insights can often be achieved[4,5].

Specific attempts to study SF in metallic nanoparticles have been performed a long ago, particularly in regards of the electronic spin susceptibility $\chi''_{spin}(\mathbf{k},\omega)$ involved in NMR $T_1$ relaxation[6, 7] and of the diamagnetic susceptibility $\chi_{dia}$[8] as well. In $T_1$ measurements the evidence of the effects related to SF is elusive, the increase of the relaxation rate for $T \rightarrow T_c^+$ being smeared out in 0D particles by rounding of the transition and the renormalization due to non linear fluctuations[7]. The SF effect on the Knight shift is also difficult to single out and again a reduction in $\chi''_{spin}(\mathbf{k=0}, \omega=0)$ in grains is observed near $T_c$[6,9].

Fluctuation diamagnetism in 0D Aluminium particles was beautifully studied in a pioneer piece of experiment by Buhrman and Halperin[8], by measuring the magnetization as a function of temperature, in constant magnetic fields. The experimental findings were analysed in terms of $\chi_{dia}$ for evanescent field, as derived in exact theories[10, 11] for the GL free energy functional in 0D condition. The value of $M_{dia}(T \approx T_c)$ at different magnetic fields were discussed by extending the zero-field equations with the replacement of the reduced temperature $\varepsilon = (T-T_c) / T_c$ by $( \varepsilon + H^2 / H_c^2 )$, with $H_c$ size-dependent critical field. The tendency towards the expected limiting slope $\chi_{dia} \propto H^{-2}$ was emphasized[8].

Recently Li et al.[12] have carried out conductivity, specific heat and susceptibility measurements in Pb nanoparticles, evidencing the quantum size effect on various SC properties. Electron tunnelling in nm-scale Al particles has been used[13] to study the structure of the electronic energy levels. The size effect on $T_c$ of lead nanoparticles embedded in amorphous matrix has been studied by Tsai et al.[14]. Gladilin et al.[15] have developed an exhaustive theory on the magnetic response and the SC properties of ultrasmall grains. For various properties of ultrasmall particles, see the reviews in Ref.16.

Here we report the results of a study of the superconducting fluctuations and of the related fluctuating diamagnetism in Lead nanoparticles. By means of high-temperature and high-field resolution SQUID measurements, the isothermal magnetization curves $M_{dia}( H, T \approx T_c^+)$ have been obtained. The magnetization curves on approaching $T_c$ from above are considered[17] to convey relevant insights on FD, in the mean-field and particularly in the critical fluctuations regime. It should be remarked that in the range of temperatures and for the size of the nanoparticles we are going to report in this paper the details of the level distribution and the effects of finite-level spacing can be neglected. In a forthcoming paper[18], by using nanoparticles having diameter $d \leq$ 100 A° the quantum size effects will be addressed.

## Experimentals

The Pb nanoparticles were produced by a modification of the Polyol Process technique [19]. All chemical reagents and solvents (Sigma-Aldrich products) had purity higher than 99%. The reaction was performed in a sealed glass vessel surmounted by a cooling column in order to obtain polyol reflux. To reach a relatively narrow distribution of the dimensions, high homogeneity of solution during the reaction process was obtained by vigorous mechanical stirring. 50 ml of tetraethylene glycol (TEG: $C_8H_{18}O_5$) were introduced in the sealed vessel and heated by a thermostatic bath at 320 °C to reach the TEG boiling. At this time a solution of PbO in TEG was introduced in the vessel. In order to produce grains at different size, the concentration of PbO in TEG and the reaction time were adjusted between 0.02 - 0.4 mol and 15 - 50 minutes, respectively.

The powder was washed several times with ethanol at the beginning and acetone at the end. To separate the fine powder fraction of Pb particles obtained at the end of the reaction process from the coarse one, an ultracentrifuge was utilized. The particles were electrically isolated each other by natural thin layers of oxide produced by ageing the particles in atmosphere for about 30 days. A Philips PW 1710 diffractometer utilizing Cu K$\alpha$ radiation ($\lambda$ = 1.5418 A°) was used for powder X-Ray Diffraction (XRD) measurements. XRD patterns between $25° < 2\theta < 65°$ were collected.

A Philips EM 208S Transmission Electron Microscope was utilized at 100 keV with a copper (Cu) grid coated by a formvar membrane for imaging the nanoparticles. Lead (Pb) nanoparticles were suspended in ethanol by sonicating, then the Cu grid was dipped in the solution. Finally, by means of a metallizer EDWARDS E306, the grids containing the Pb nanoparticles were coated by a very thin carbon layer in order to avoid the membrane breaking.

The nanoparticles have also been analyzed through Atomic Force Microscopy (AFM). The AFM (Digital Instruments Nanoscope IIIa) was equipped with a sharpened silicon tip with a radius less than 5 nm. The images of the surface profiles were obtained by operating the AFM in the tapping mode, with a scan size and rate of 2 μm and 2 Hz, respectively.

The X-Ray diffraction peaks in Fig1a) are from fcc Pb, with a smaller peak corresponding to the thin layer of superficial oxide, that can be estimated around 5% in weight. In the same Figure (part b and part c) TEM images are reported . From them the oxide thickness can be estimated to an extent of about 10% of the nanoparticle diameter.

The particles size distribution, based on the TEM images, fitted to a log-normal distribution of the particle diameters, d, is shown in Fig. 1(d), for sample (1). In Fig.2 the AFM image for sample (3) and the related particles size distribution, fitted to a log-normal distribution of the particle diameters, d, are reported.

The expected diameter $d_e = \exp(\mu + \sigma^2/2)$, the standard deviation of the particle diameters, $\sigma_d = \exp(\mu + \sigma^2/2)\ (\exp(\sigma^2) - 1)^{1/2}$, where $\mu$ and $\sigma$ are the mean and standard deviation of ln$d$, respectively, treated as fitting variables, and the median diameter $d_m$ , obtained by finding the value at which the integral of the distribution given by a log-normal function is equal on each side of $d_m$, are reported in Table I. The median diameter is taken as the representative size for further analysis.

**Table I**. Expected (average) diameters of the particles, standard deviation of the particle diameters, median diameters of the particles, superconducting transition temperatures, limits of the critical region (see text) and critical fields of the grains (as obtained from the extrapolation of $T_c$ (H) for relatively low field according to Eq.2). It is noted that by using an effective, properly averaged, diameter of the grains the estimates of $\varepsilon_c$ do not change much. The sample (2b) has been kindly provided by W.-H. Li, National Central University of Chung-Li (Taiwan).

| | (1) | (2a) | (2b) | (3) |
|---|---|---|---|---|
| | $d_{e1}$ = 160 Å | $d_2$=250 Å | $d_2 \approx$ 400 Å | $d_{e3}$ = 750 Å |
| | $\sigma_{d1}$ = 40 Å | $\delta d_2$=60 Å | ------- | $\sigma_{d3}$ = 200 Å |
| | $dm_1$ = 170 Å | ----------- | ---------- | $dm_3$ = 720 Å |
| $T_c (H\to 0)$/ (K) | 7.09 ± 0.005 | 7.09 ± 0.005 | 7.1 ± 0.03 | 7.09± 0.005 |
| $\varepsilon_c$ | $\approx 9\ 10^{-2}$ | $\approx 2.2\ 10^{-2}$ | ------- | $\approx 1.1\ 10^{-2}$ |
| $H_c^{grain}$ (Oe) | 2500±400 | ------- | | 1150±100 |

From the thickness of the oxide layers, by assuming the condition of random loose packing a density of nanoparticles approximately 1/3 with respect to bulk Pb could be inferred.
The zero-field transition temperatures (see Table I) have been obtained by extrapolating linearly to zero the susceptibility in a field of 1 Oe plotted vs $T^4$ for $T\to T_c^-$ (See inset in Fig.3(a) ).

In Figs. 3 the blow-up of the data for $\chi_{dia}$, for evanescent field, around $T_c$ are reported as a function of the reduced temperature $\varepsilon = ( T - T_c ) / T_c$ , for sample (1) and sample (3).

The field dependence of the transition temperature $T_c(H)$ has been estimated by extrapolating to zero the data for ( $M_{dia}$ /H ) obtained in the temperature range where this quantity varies linearly with $T^4$ ( Fig.4).

In Figs. 5 representative isothermal magnetization curves are reported in the temperature ranges around $T_c(0)$. The values of $M_{dia}$ due to FD have been obtained by subtracting from the raw data the magnetization measured at temperature ( around 8°K), where the SF are negligible and only a paramagnetic contribution was found to be present.

## Analysis of the data

The temperature range $\varepsilon < \varepsilon_c$ where critical, non-mean field fluctuations are expected to occur can be estimated according to slightly different criteria. The Ginzburg-Levanyuk criterium in 0D and in the assumption of the BCS condition for $\xi_0$ yields[1] $\varepsilon_c \approx 13.3\ ( T_c / T_F ) ( \xi_0^3/v )^{1/2}$, where $T_F$ is the Fermi temperature and v the volume of the particle.
The critical reduced temperature can be defined as the one at which the first-order fluctuation correction to the mean field behaviour of $<|\psi|^2>$ is abandoned [11], yielding
$\varepsilon_c \approx 0.95\ [ N(0)\ v\ k_B\ T_c ]^{-1/2}$ , N(0) being the single–spin density of states per unit volume. By assuming that the electron mean free path is limited by surface scattering one could use[8]
$\varepsilon_c \approx (6\ k_B\ T_c)^{1/2} / (d/2)^{3/2} |\ T_c\ dH_c/dT\ |_{Tc}$ .
For lead $T_F$ = 1.1.$10^5$ K, the electron density is n= 1.32 .$10^{23}$ $cm^{-3}$ and $\xi_0$ = 900 Å. The experimental result for the thermodynamic bulk critical field yields $| dH_c/dT |_{Tc}$ = 81 Oe / °K.
The critical reduced temperatures $\varepsilon_c$ reported in Table I are average values of the estimates according to the criteria recalled above.

In the framework of the mean-field GL theory, for $\varepsilon > \varepsilon_c$ and $H << H_c$ the single particle magnetization is[1]

$$M_{dia}= - k_BT\ H\ (4\pi^2\ \xi_0^2 d^2 / 5\ \Phi_0^2)/[\varepsilon + (2\pi^2\ \xi_0^2 d^2\ H^2 /5\ \Phi_0^2)]=$$

$$= - k_B T\, H\, 2\, (H_c^{grain})^{-2} / [\varepsilon + H^2/(H_c^{grain})^2] \qquad [1$$

In this Eq. $H_c^{grain} = (2.5)^{1/2}\, \Phi_0 / \pi\, \xi_0 d$ can be defined as zero-temperature critical field of the grain. In fact in the same theoretical framework the field dependence of the transition temperature is given by

$$T_c(H) = T_c(0)\, [1- (4\pi^2 \xi_0^2\, H^2\, d^2/10\Phi_0^2)] \equiv T_c(0)\, [1- H^2 / (H_c^{grain})^2]. \qquad [2$$

It is noted that Eq. 1 implies an upturn in the field dependence of $M_{dia}$ around the field

$$H_{up} = \varepsilon^{1/2}\, (2.5)^{1/2}\, \Phi_0/ \pi\, \xi_0 d \equiv \varepsilon^{1/2}\, H_c^{grain} \qquad [3$$

The solid lines in Fig. 5(a) correspond to Eq. 1, with no free parameters, having used for $H_c^{grain}$ the value 1150 Oe , as indicated from the field dependence of the transition temperatures on the basis Eq.2. The experimental data for $H_{up}$ and $H_c^{grain}$ are not far from the ones expected from the above Eqs., for $\xi_0$=900 Å, particularly if an effective grain size larger than the nominal one is taken into account because of the size distribution. For instance, for the sample at d=750 Å Eqs. 2 and 3 yield $H_{up} \approx 1.5 \cdot 10^3\ \varepsilon^{1/2}$ Oe and $H_c^{grain} \approx 1400$ Oe. It can be remarked that the distribution in d is particularly detrimental for small size of the grains and makes the estimate of the field dependence of $T_c$ for strong H affected by large errors ( see Fig.4). The zero-temperature critical fields for grains where $H_c^{grain}$ $(T\to 0)$ is much larger than in the bulk material can hardly be estimated by extrapolating the data for $T_c(H)$ according to Eq.2.

For sample (3) (d=750 Å) the comparison of the field dependence of $T_c$ ( Fig.4) with the curve derived on the basis of the temperature dependence of $H_c$ [12] ( for the sample at d =860 Å) indicates that Eq.2 is of valuable validity, but only for $H \le H_c^{grain}/2$. This limit of validity could be expected in the light of the general structure of the theory, based on the first-order fluctuation correction.

From Fig.5(a) it appears that the experimental data at T = 7.16 K are very well fitted by Eq.1. On the contrary, for temperatures closer to $T_c$, and particularly for T= 7.095 K, well inside the critical region, the departure of $M_{dia}$ from the behaviour expected on the basis of Eq.1 is noticeable. We shall see that the magnetization curves derived from the full form of the GL functional and of the exact partition function will account for this departure. From Fig.5(a) one can remark that the isothermal magnetization curves convey information on the FD much more reliable of the isofield data as a function of temperature (solid circles), commonly used.

Therefore the general conclusion is that above the critical region the experimental findings of the fluctuating diamagnetism follow rather well the theoretical predictions outlined above, the first-order fluctuation correction to the GL functional being basically appropriate and the parameters scaling in the expected form. The departures of the experimental results for $M_{dia}$ with respect to the magnetization given in Eq.1 appearing in Fig.5(a) are just the signature that when entering in the critical region the susceptibility is smaller than the one expected in the framework of the first order fluctuations correction (See Eq.1 and the solid lines in Fig.3).

Now we turn to the discussion of the magnetization curves in the critical region. It is noted that for sample (1) all the measurements practically refer to $\varepsilon \le \varepsilon_c$ , where the term in $|\Psi|^4$ in the GL functional cannot be disregarded. In the critical region, for $H \ll H_{up}$ the magnetization is still linear in the field and the susceptibility can be written

$$\chi_{dia} \approx - d^{3/2}\, (12\, k_B T_c)^{1/2} / 33.9\, \lambda_L(0)\, \Phi_0 \approx - d^{3/2}\ 0.6\, (k_B T_c)^{1/2}/\Phi_0^2\, H_c^{bulk} \qquad [4$$

with weak temperature dependence (see plots in Ref.11). With the value $H_c^{bulk}$ =800 Oe for the bulk critical field, Eq.4 yields $\chi_{dia}$ ( $T \approx T_c$ ) = -1.2 . $10^{-5}$ for the sample at d = 750 Å, in good agreement with the data in Figs.3. The scaling factor of $\chi_{dia}$ ( $T \approx T_c$ ) with the grain size is not exactly $d^{3/2}$ and this is likely to be the trivial consequence of the approximation in the estimate of the filling factor accounting for the presence of the insulating oxide.

From the comparison of the solid lines with the experimental data in Fig.5 and from the temperature dependence of the upturn field (Fig.6) it is evident that the mean field correction breaks down below about 7.13 °K in sample (3) and in almost the whole temperature range that has been explored in sample (1).

In the critical region the role of the magnetic field can be discussed starting from the exact expression of the GL functional and by using for the partition function the expression[1]

$$Z_{(0)} = [\ \pi^3 v\ k_B T / 2b\ ]^{1/2} \quad \exp(x^2)\ (1 - \mathrm{erf}(x)) \qquad [5$$

with $x = a(H)\ (v / 2\ b\ k_B\ T)^{1/2}$

b being the coefficient of the $|\ \psi\ |^4$ term while a is the coefficient of the term in $|\psi|^2$ including $H^2 d^2$ /10 . It is recalled that the first-order fluctuation corrections corresponds to the approximation of the free energy in the form $F_{(0)} = -\ k_B T \ln(\pi/\alpha\varepsilon)$ , as it has been used for $\varepsilon > \varepsilon_c$ .

From Eq.5 $M_{dia}$ vs H has been derived. The parameter a becomes $\alpha\ \ k_B\ T_c[\varepsilon + (\ H/H_c^{grain})^2\ ]$, while the factor $(v\ /2bk_B\ T\ )^{1/2}\ \alpha\ k_B\ T_c$ has to be estimated in correspondence to the relation $(\alpha^2/b) =\ 8\pi^2\ N(0) / 7\ \zeta\ (3)$ and $N(0)\ v\ k_B\ T_c \approx 0.95\ \varepsilon_c$ .

The magnetization (per unit volume) turns out

$$M_{dia} =\ (2\alpha/\sqrt{b}\sqrt{v})\ k_B^{3/2} T_c T^{1/2}\ \ [H/\ (H_c^{grain}\ )^2]\ [x - e^{-x^2} / \sqrt{\pi}(1-\mathrm{erf}x)] \qquad [6$$

and $H_{up}$ can be estimated by numerical procedure. Eq.6, in the limit $H \to 0$ yields the susceptibility including the quartic term in the GL functional.

In Fig. 7 the curves $M_{dia}$ vs H are reported for representative temperatures. It is noted that for sample (3) at T=7.16 K the curve coincides with the one given in Fig.5(a), as expected since at this temperature $\varepsilon > \varepsilon_c$. At variance, for the other two temperatures significant differences appear with respect to the ones in Fig.5(a). Using the complete form of the partition function the magnetization curves fit rather well the experimental findings, particularly for $M_{dia}$ ( $H \approx H_{up}$), having kept all the parameters unchanged. Thus the role of the term $|\Psi|^4$ in the GL functional is emphasized and quantitatively taken into account.

In part (b) of Fig. 7 the theoretical curves derived for the sample at d=160 A° are reported, without direct attempt to fit the experimental data. From the comparison with the magnetization curves in Fig.5(b) one can observe that the trend of $M_{dia}$ vs H is rather well reproduced up to a field strength of the order of $H_{up}$ , in the derivation having used for the critical field of the grain the value $H_c^{grain}$ = 2500 Oe (Table I) . We remark that for small grain size the theoretical curves are very sensitive to the value of $H_c^{grain}$, involved through the factor a in Eq.5, difficult to be estimated with good precision by means of Eq.3 on the basis of the field dependence of $T_c$. Still the value of the upturn field and the magnetization for $H=H_{up}$ are rather well justified. For relatively large field the experimental data appear to decrease towards zero with H faster than the theoretical predictions. An attempt to take into account the size distribution of the grains reported in Fig.1(d) does not lead to significant improvement in regards of the behaviour for strong fields. On the other hand one should observe that at strong field the experimental error increases rapidly, in view of the subtraction procedure which involves a quantity which decreases on increasing H ($M_{dia}$) while the paramagnetic term is increasing. Furthermore one should take into account that for grains at diameter smaller than about 80-100 Å the quantum size effects could drastically modify the superconducting properties [16, 1]. Attempts to obtain more selective samples at narrow distribution in diameters of the grains are under way[18].

From the behaviour of $M_{dia}$ in the critical region reported in Fig.7(b) one can note that the upturn field increases on increasing temperature and that the diamagnetic magnetization $M_{dia}$ ($H=H_{up}$) increases when $H_{up}$ decreases. This means that qualitatively the behaviour predicted for the magnetization curves in the first order correction theory is still valid even in the critical region, although the temperature behaviour of the susceptibility and of the upturn field is strongly modified. This is somewhat reflected in Fig.6, since by scaling the reduced temperature in term of $\varepsilon_c$ for each size, the quantity ($H_{up}$ d ) approximately keeps a size-independent value, in spite of the breakdown of the mean field result $H_{up} \propto \varepsilon^{1/2}/d$. The inset in Fig.6 reports the temperature behaviour of $H_{up}$ (normalized to the value slightly above $\varepsilon_c$ ) expected in the critical region according to our derivation based on the exact GL functional and the full form of the partition function. The experimental findings for ($H_{up}$ d ) vs $\varepsilon/\varepsilon_c$ are rather well accounted for.

## Summarizing remarks

By means of magnetization measurements we have studied the superconducting fluctuations and the related fluctuation diamagnetism above the superconducting transition temperature in Pb nanoparticles of size smaller than the coherence length. The isothermal field dependence of the diamagnetic magnetization $M_{dia}$ above $T_c$ has been discussed in the framework of exact theories based on the Ginzburg-Landau functional in the zero-dimensional condition.

The first-order fluctuation correction has been confirmed a suitable approximation only outside the critical region and it describes rather well the behaviour $M_{dia}$ in finite field H not too close to the critical field. Also the scaling properties of $dT_c(H)/dH$ for small field and of the upturn field $H_{up}$ in the magnetization curves are well justified by that approximation, for $\varepsilon>\varepsilon_c$.

In the critical region the role of the field and the limits of validity of the mean-field fluctuation regime have been analysed by comparing the experimental findings to the derivation of $M_{dia}$ as a function of the magnetic field carried out starting from the complete form of the GL functional and with the exact expression of the zero-dimensional partition function. In this way, for the sample at average diameter of the grains around 750 Å the fluctuating diamagnetism can be well justified even in the temperature range where the role of the $|\Psi|^4$ term in the GL functional is crucial, without any adjustable parameter. For the sample at smallest average diameter, around 160A°, the agreement of the numerically derived behaviours of $M_{dia}$ with the experimental findings is again good for field of the order of $H_{up}$ . Poor agreement between the theoretically predicted $M_{dia}$ vs H and the experimental trend is observed for fields above $H_{up}$ , when the fluctuating diamagnetic contribution is going towards zero and the subtraction procedure of the paramagnetic term implies large errors. Possible effects related to the smallest grains sizes can be suspected.

The temperature dependence of the upturn field and the scaling properties with the grain size appear satisfactorily verified outside as well as inside the critical region, the product ($H_{up}$ d) vs reduced temperature approximately keeping a size-independent value and following the expected temperature behaviour, even though the mean field result $H_{up} \propto ( \varepsilon^{1/2}/d )$ evidently breaks down.

The relevance of the magnetization curves for non-evanescent field and of the upturn field $H_{up}$ for the study of the fluctuating diamagnetism above the superconducting transition temperature has been emphasized.

## Acknowledgments

Preliminary measurements carried out by I. Zucca are gratefully acknowledged. Prof. W.-H. Li , ( National Central University of Chung-Li ) is thanked for the loan of a batch of Pb nanoparticles used for comparison. A.A. Varlamov is gratefully thanked for stimulating discussions and for having introduced the authors to the study of the superconducting fluctuations. The authors thank the Centre CISME (Centro Interdipartimentale di Servizio per la Microscopia Elettronica) of the "Federico II" University for access to TEM device. Useful discussions with P. Carretta are gratefully acknowledged.

The work has been carried out in the framework of a FIRB-MIUR project "microsistemi basati su materiali magnetici innovativi strutturati su scala nanoscopica." (Pnr 2001-2003).

## Captions for Figures

**Fig. 1** (a) XRD of Pb nano-powder. (b) TEM images of Pb nanoparticles (Mag. 40000 X). (c) TEM images of Pb nanoparticles (Mag. 350000 X). (d) Size histogram of Pb nanoparticles produced under experimental conditions shown in (b). The curve in the part d represents a log-normal distribution, which has a median diameter $d_{m1}$ = 170 Å (sample (1)) and a standard deviation of the particle diameters $\sigma_{d1}$ = 40 Å.

**Fig. 2** (a) AFM image of Pb powder onto mica substrate obtained by increasing the concentration of PbO in TEG and the reaction time up to threefold. (b) Size histogram of Pb particles produced under experimental conditions shown in (a). The curve in the part (d) represents a log-normal distribution, which has a median diameter $d_{m3}$ = 720 Å (sample (3)) and a standard deviation of the particle diameters $\sigma_{d3}$ = 200 Å.

**Fig.3** Blow-up of the temperature dependence of the susceptibility, in the limit of evanescent magnetic field, around $T_c$ in sample (3) (a) and in sample (1) (b). The solid circles are the data from the isothermal magnetization curves while the empty squares are from the isofield measurements as a function of temperature. The solid lines track the behaviour of $\chi_{dia}$ in the assumption that the non-linear fluctuations can be neglected (mean field fluctuations regime). The inset in Fig.3(a) shows how the transition temperature has been determined. $\varepsilon$ is the reduced temperature .

**Fig.4** Field dependence of the superconducting transition temperatures for sample (3) ( • ) and for sample (1) ( ∇ ). The dotted lines are the behaviours according to $H_c(T) = H_c(0)\,[\,1- (T/T_c)^{\alpha}\,]$, with $\alpha$= 2.15 and 2.2, in correspondence to the critical field 1200 Oe ( for a sample at d=860 Å) and for the indicative value $H_c^{grain}$ (0)= 5000 Oe for d =160 Å (see Ref.12 ). The solid lines corresponds to Eq.2 in the text and it appears to hold for $H \leq H_c^{grain}(0)/2$. For sample (1) the extrapolation yields $H_c^{grain}$ around 2500 Oe ( see Table I).

**Fig.5** (a) Magnetization $M_{dia}$ vs. H in sample (3) at representative temperatures above $T_c$. The solid lines correspond to Eq.1 in the text for critical field of the grain 1150 Oe . Similar curves have been obtained for samples (2). For $\varepsilon \leq \varepsilon_c$ the curves depart from the behaviour described by Eq.1.
(b) Magnetization curves for sample (1), all corresponding to temperature range where $\varepsilon < \varepsilon_c$ , namely within the critical region. The open circles in part (a) correspond to the data obtained from the isofield measurements as a function of temperature, with large experimental errors.

**Fig. 6** Temperature behaviour of the upturn field $H_{up}$, normalized to the grain size, as a function of the reduced temperature ( • sample 3; ♦ sample 1; □ and ∇ samples 2) . The upturn field appears to scale rather well with the grain size on the whole temperature range, while the break down of the first order fluctuation correction (Eq.3) occurs for $\varepsilon \leq \varepsilon_c$. The inset shows the temperature behaviour of $H_{up}$ (normalized to the value slightly above the critical temperature) as obtained from the exact expression of the GL functional and the full form of the partition function (see text).

**Fig. 7** Magnetization curves at different representative temperatures derived as explained in the text from the exact GL functional and the complete form of the zero-dimensional partition function. Part (a) of the figure refers to the sample at average diameter of the grains 750 Å, with the experimental data taken from Fig.5 (a). Part (b) reports the theoretical magnetization curves expected at representative temperatures for sample (1), at average diameter size of 160 Å.

**Fig. 1**

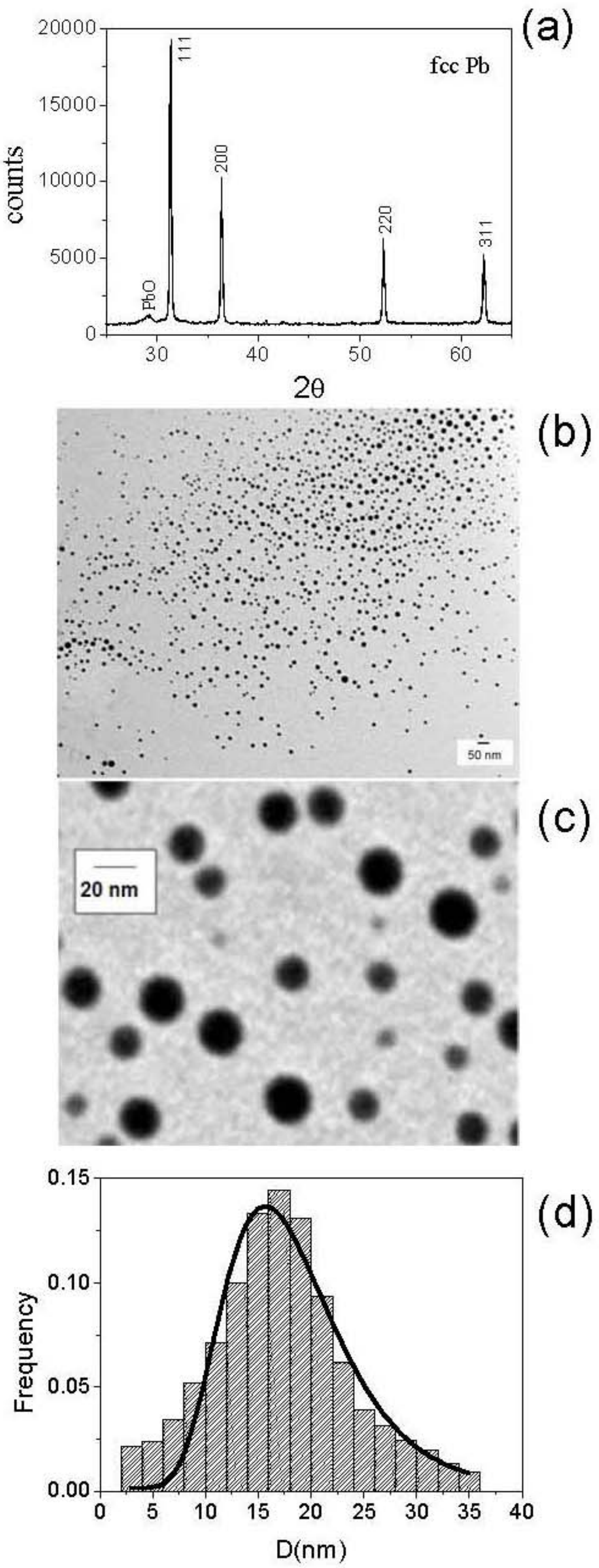

**Fig. 2**

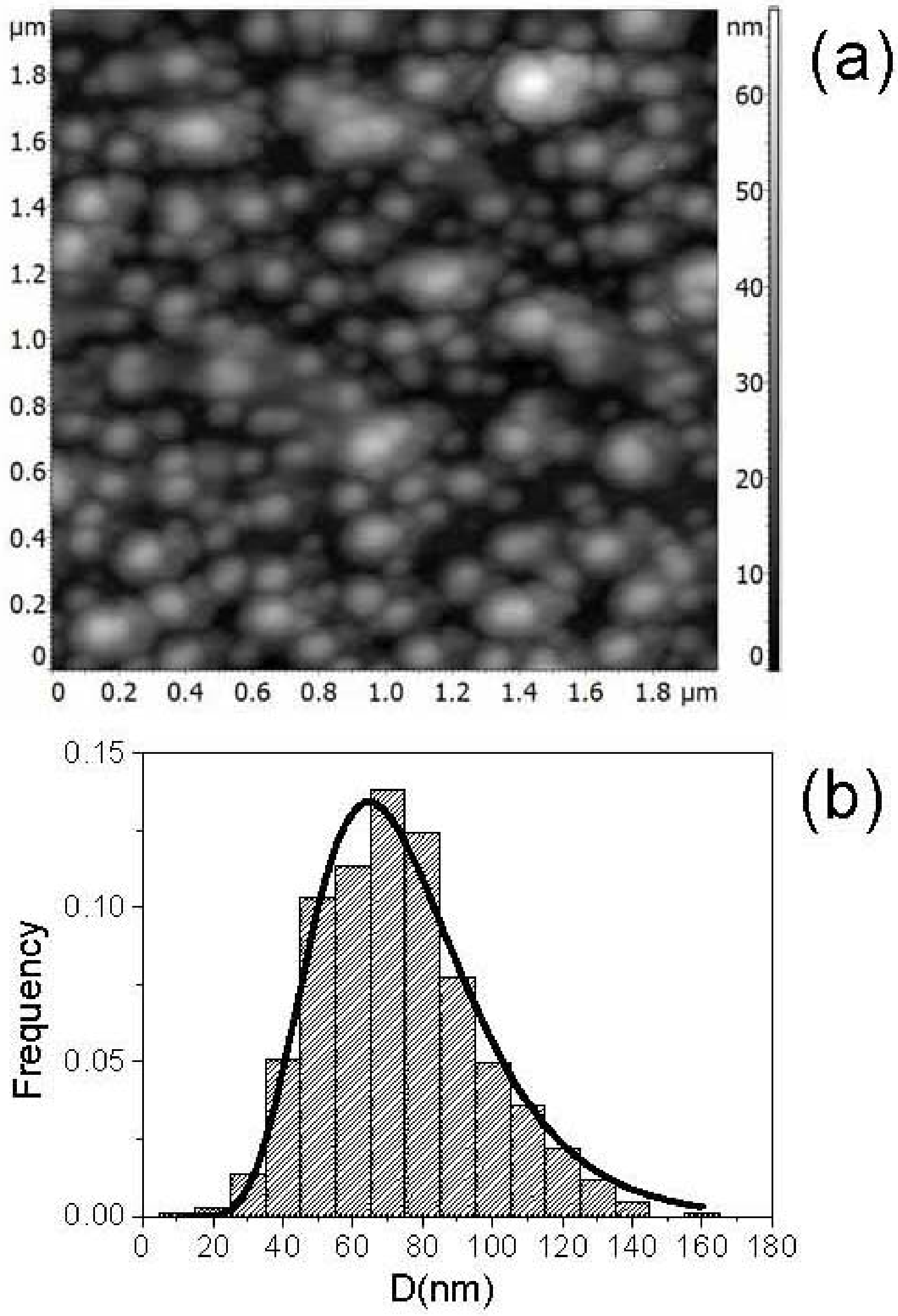

μm
1.8
1.6
1.4
1.2
1.0
0.8
0.6
0.4
0.2
0
0 0.2 0.4 0.6 0.8 1.0 1.2 1.4 1.6 1.8 μm
nm
60
50
40
30
20
10
0
(a)
0.15
0.10
0.05
0.00
Frequency
0 20 40 60 80 100 120 140 160 180
D(nm)
(b)

Fig. 3a

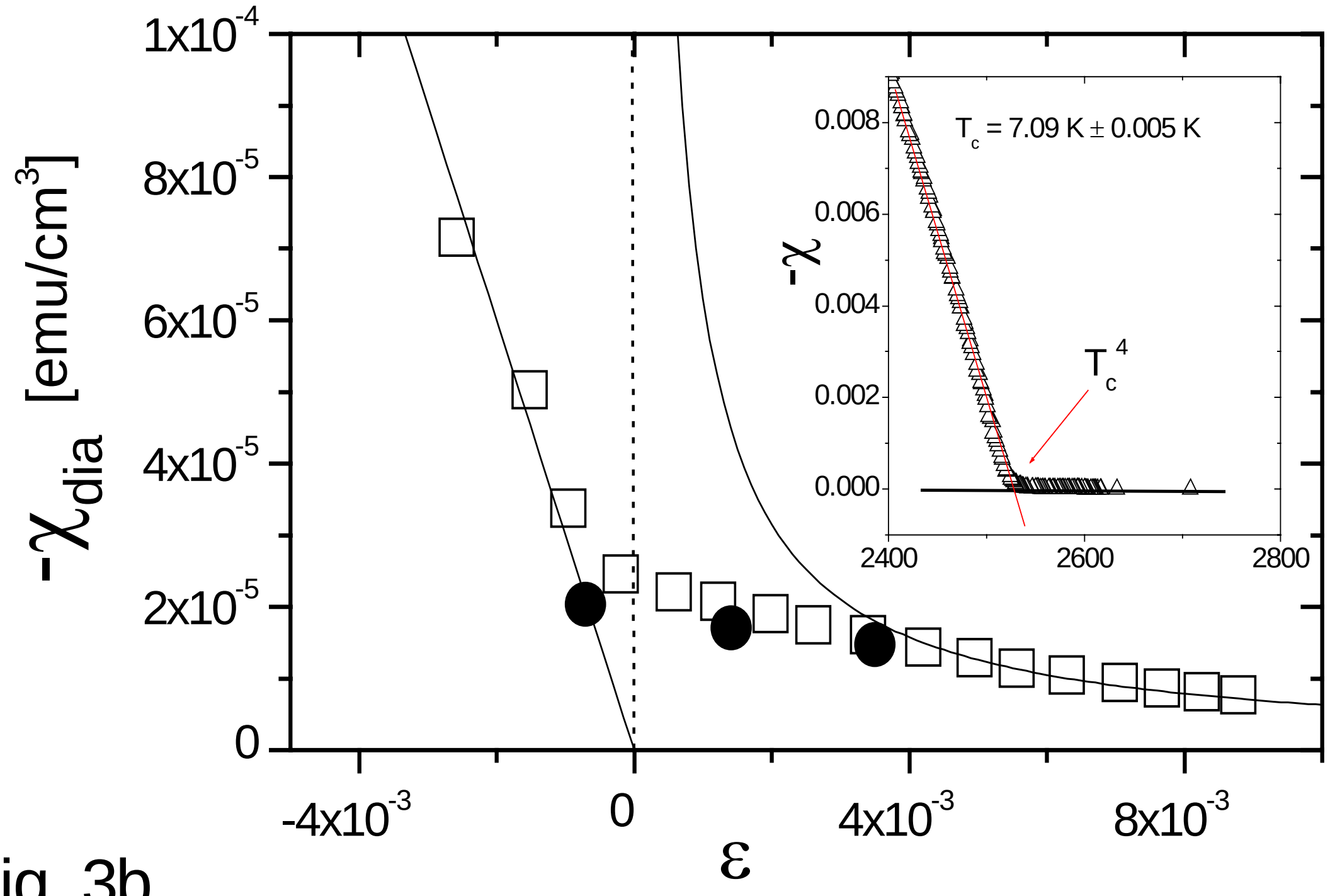

Fig. 3b

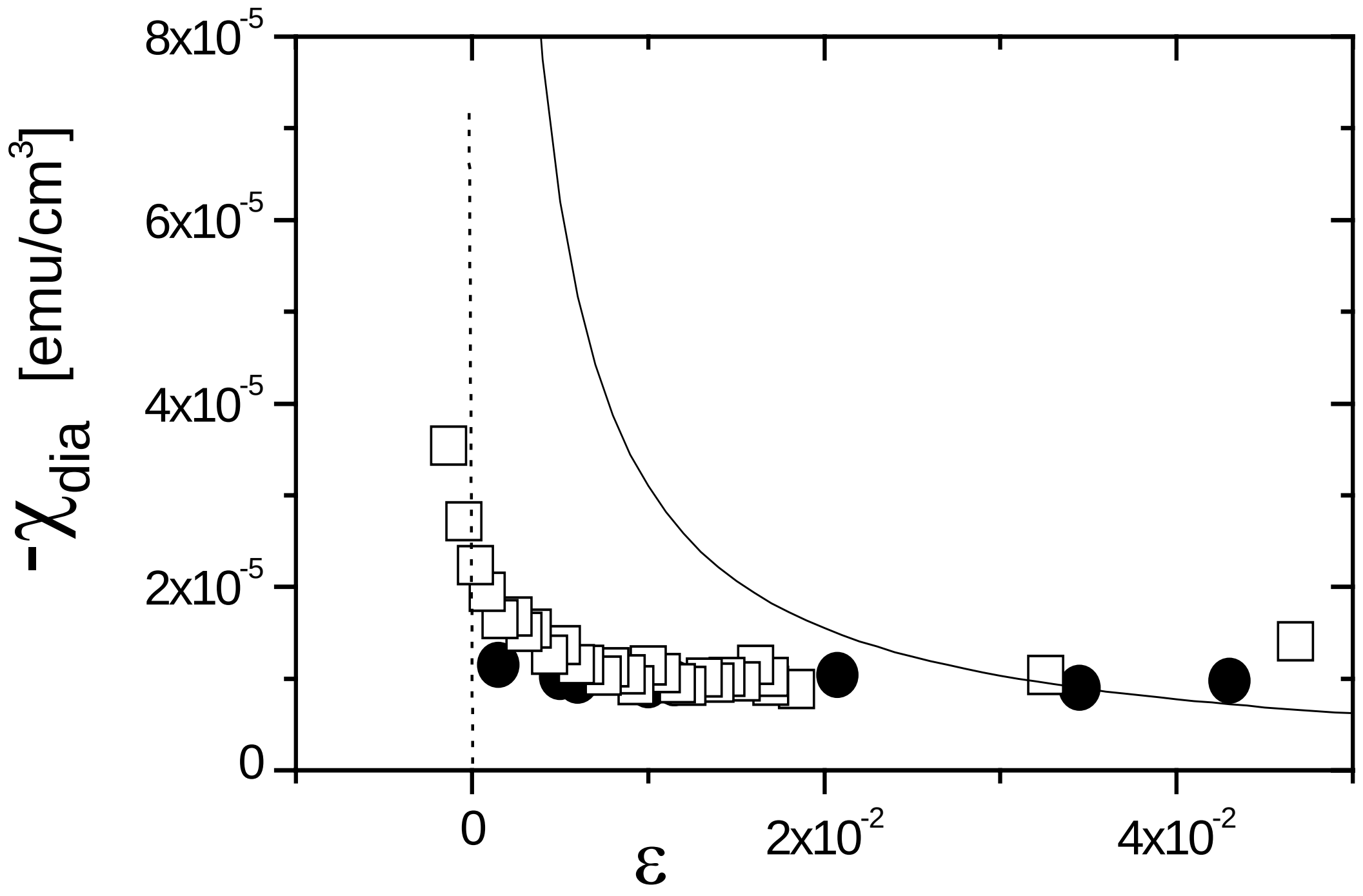

Fig.4

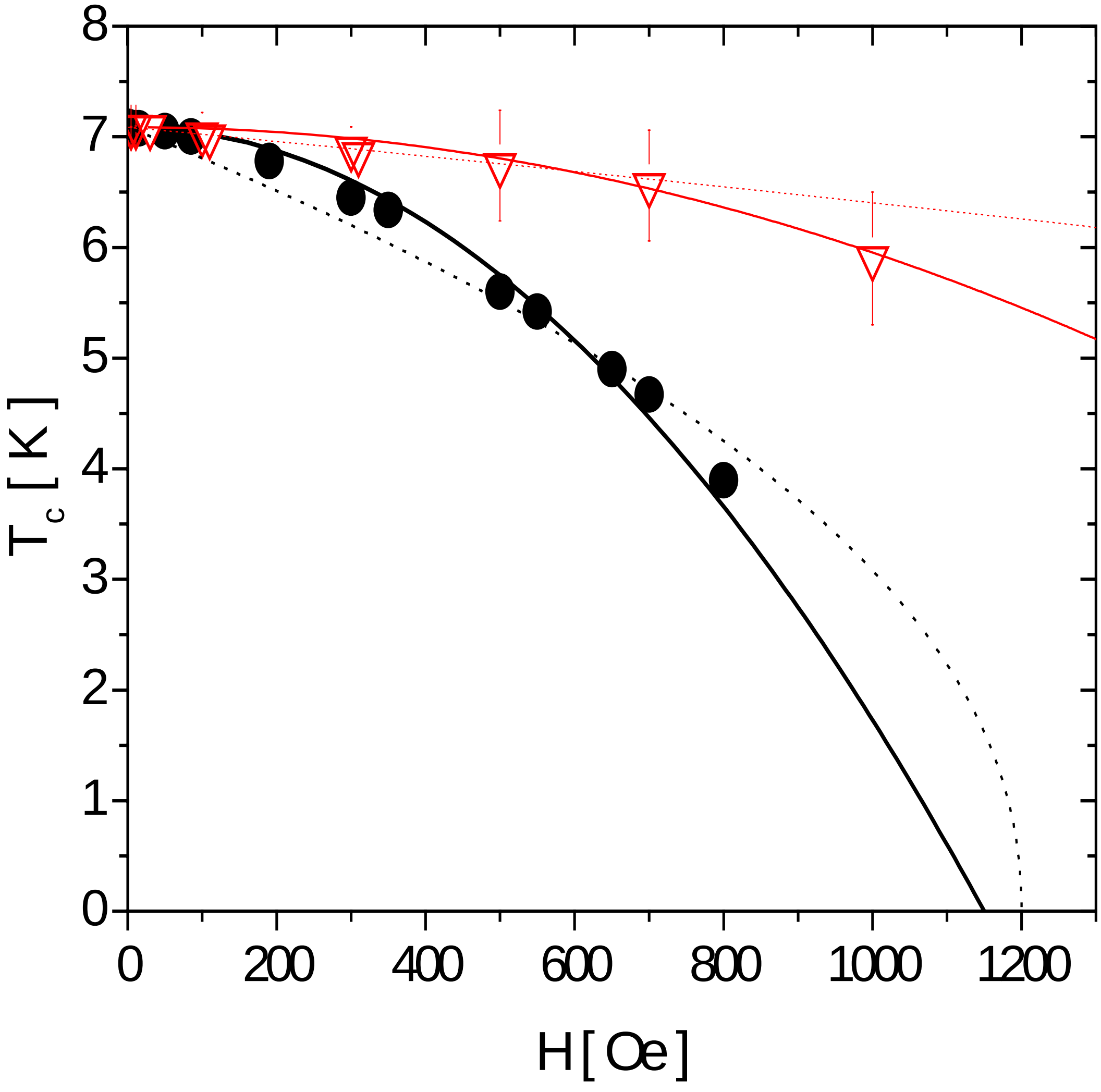

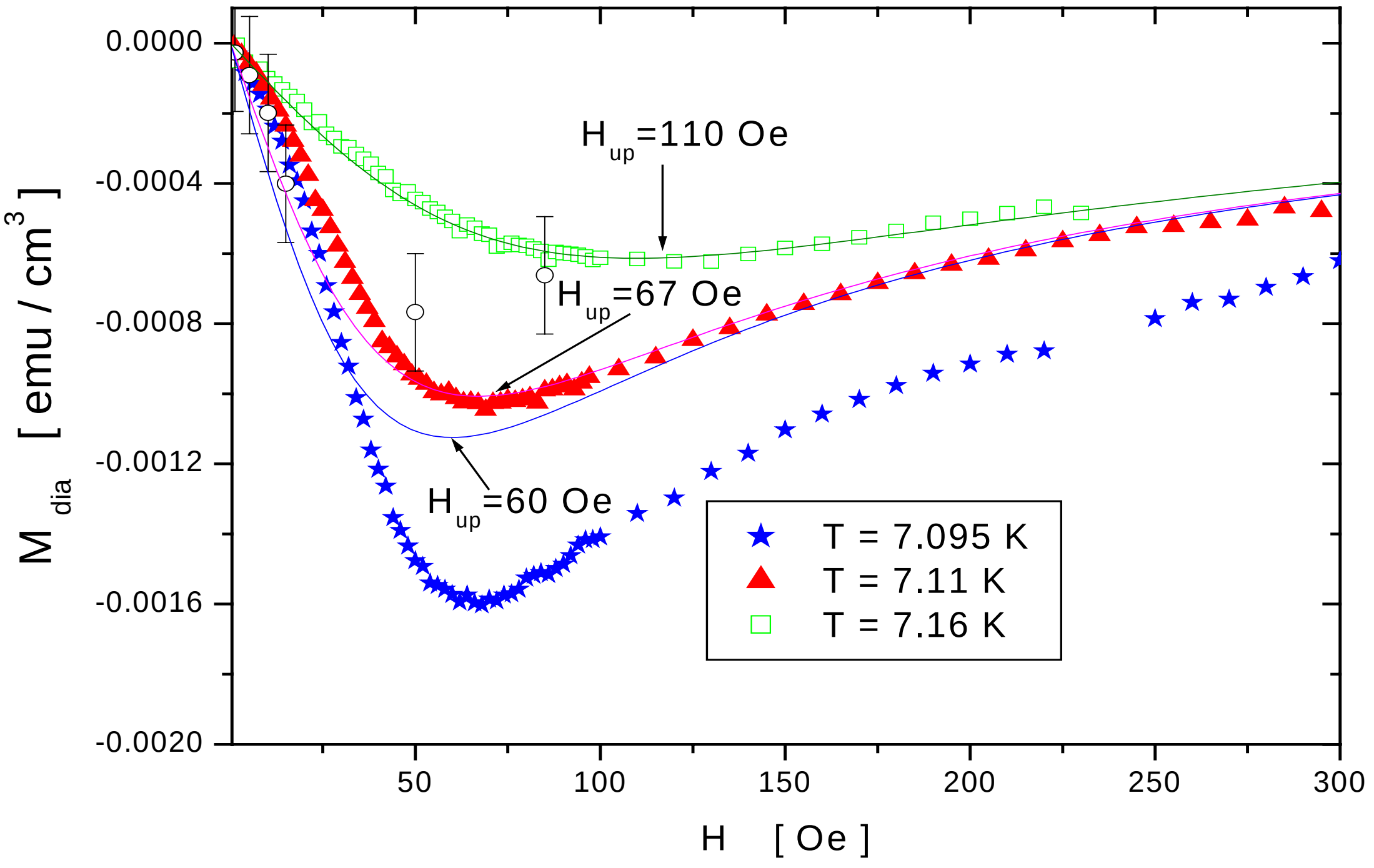

0.0000
-0.0004
-0.0008
-0.0012
-0.0016
-0.0020
50
100
150
200
250
300
H [ Oe ]
M dia [ emu / cm 3 ]
H up =110 Oe
H up =67 Oe
H up =60 Oe
T = 7.095 K
T = 7.11 K
T = 7.16 K

Fig.5(a)

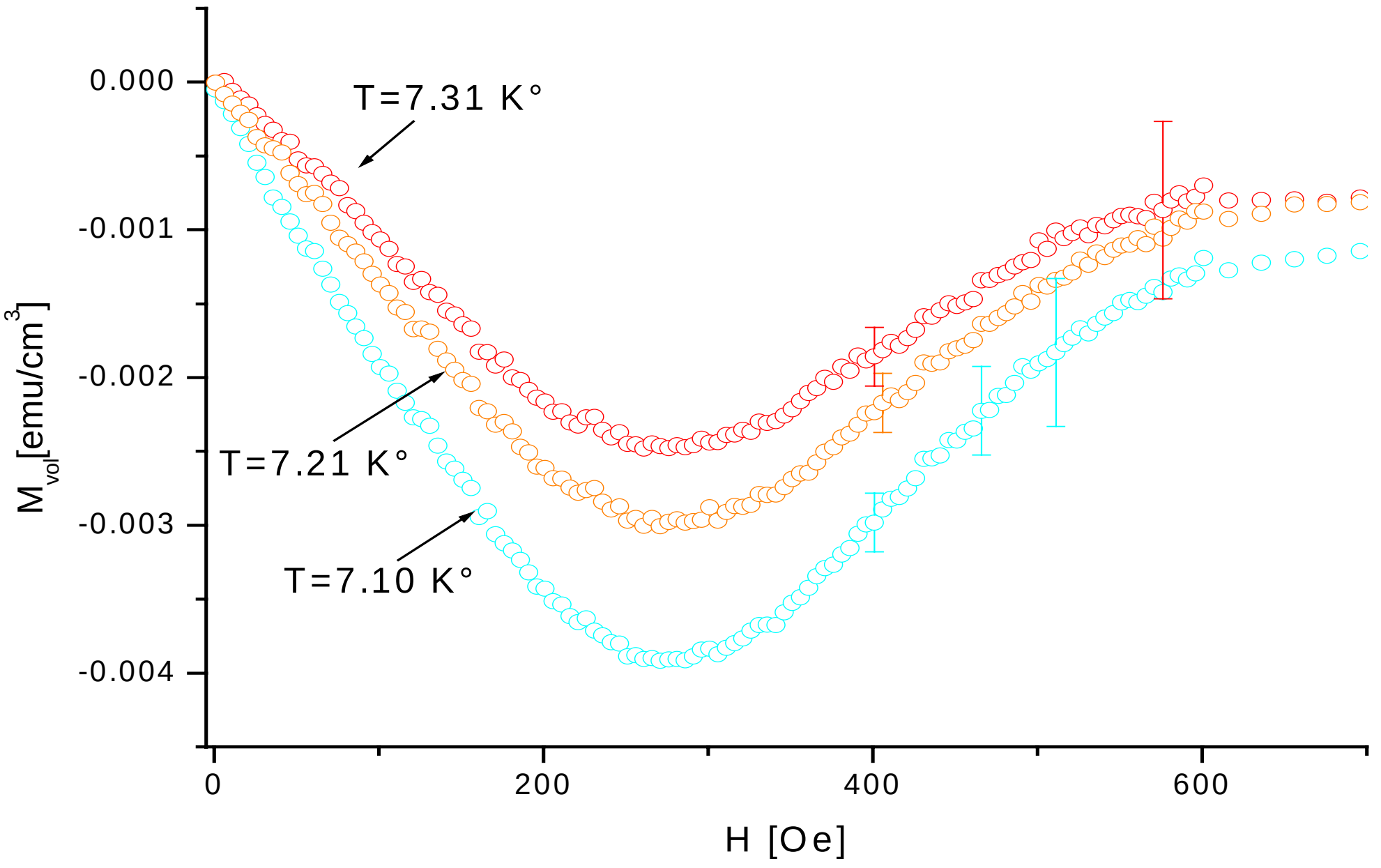

Fig.5(b)

Fig.6

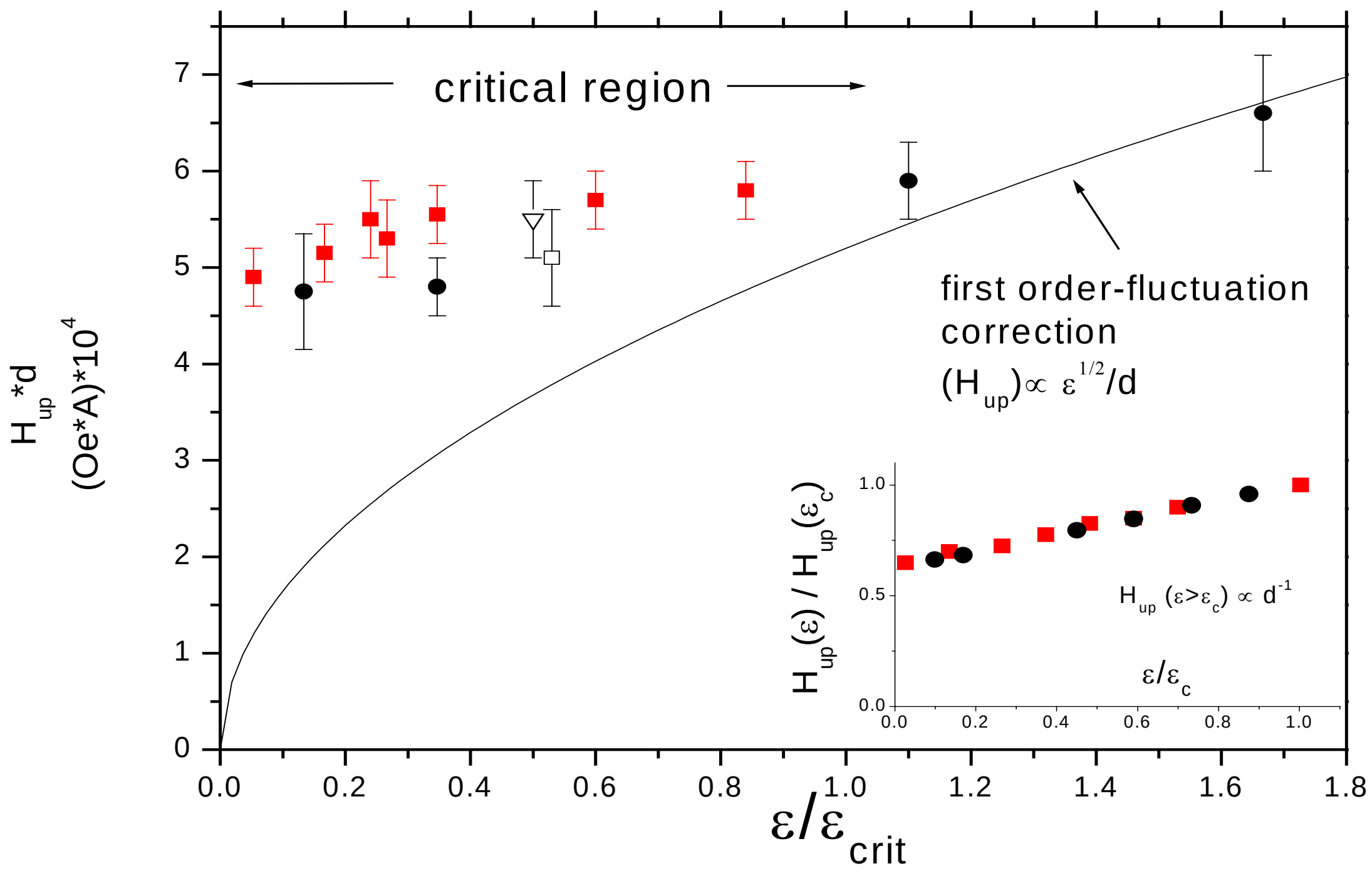

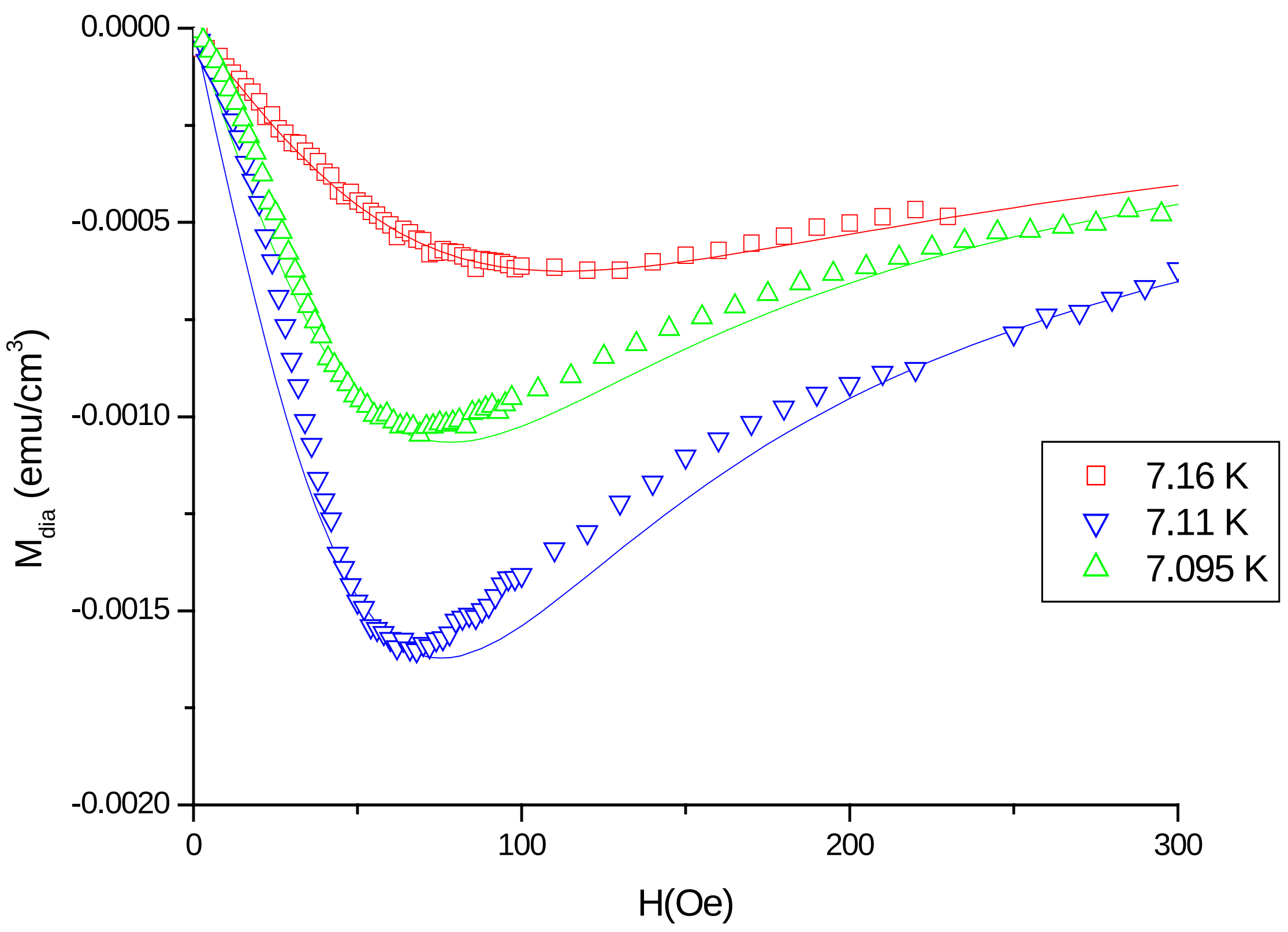

Fig.7(a)

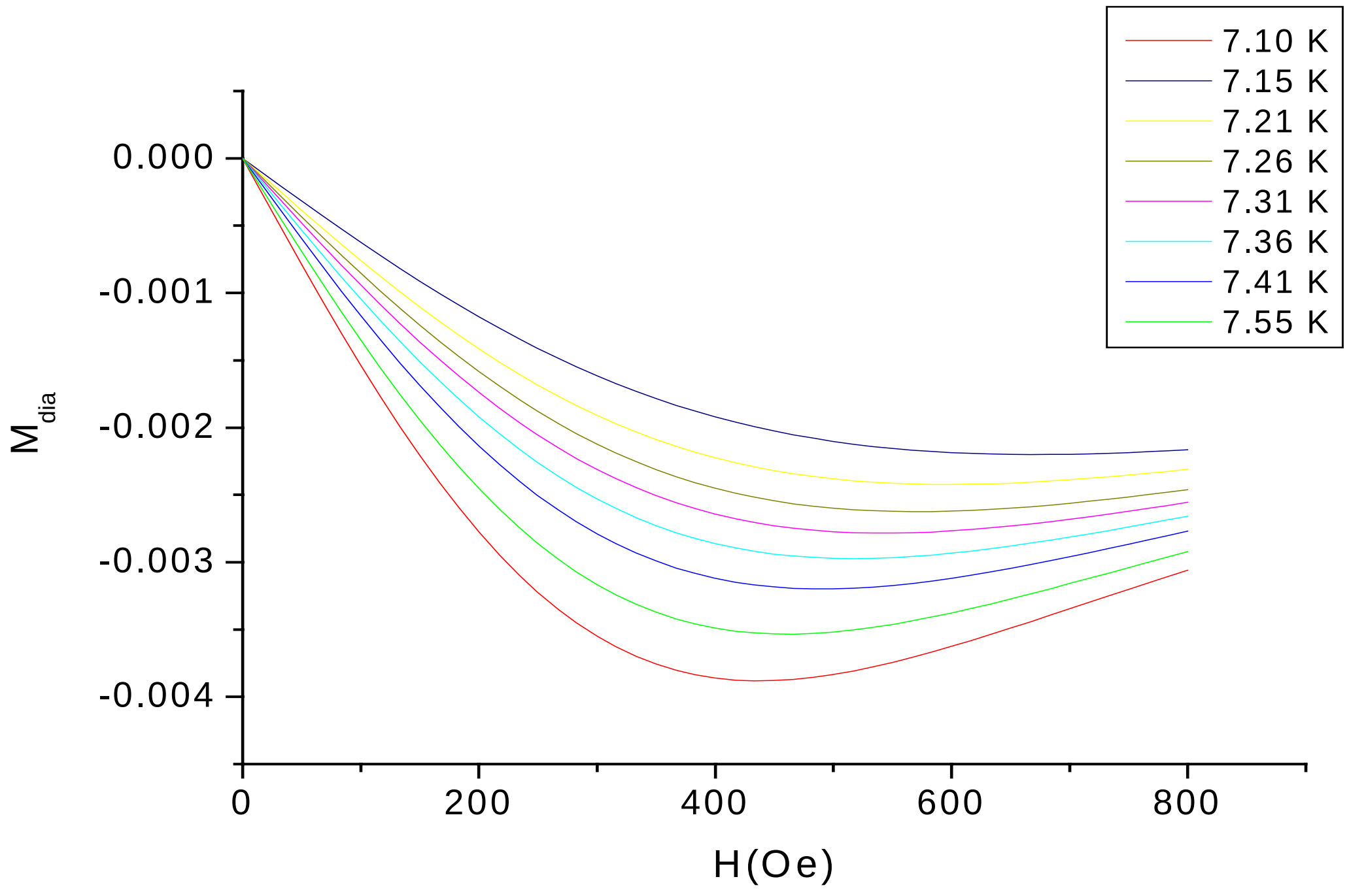

Fig. 7(b)